\documentclass[aps,a4paper,twocolumn]{revtex4-2}
\usepackage{amsmath}
\usepackage{graphicx}
\usepackage{subfigure}
\usepackage[usenames,dvipsnames]{color}
\definecolor{darkblue}{RGB}{0,0,196}
\usepackage[colorlinks=true,linkcolor=darkblue,citecolor=darkblue,urlcolor=darkblue]{hyperref}
\usepackage{setspace}
\usepackage{multirow}
\usepackage{hyperref}
\usepackage{xcolor}
\hypersetup{
  colorlinks   = true, 
  urlcolor     = red, 
  linkcolor    = blue, 
  citecolor   = blue 
}
\usepackage{graphicx}
\usepackage{amsmath,bbm}
\usepackage{amssymb,bm}
\usepackage{yfonts}
\usepackage{comment}
\usepackage{tabularx}
\usepackage{xspace}

\usepackage{booktabs} 
\usepackage{float}    
\usepackage{placeins} 

\usepackage[usenames,dvipsnames]{color}
\definecolor{darkblue}{RGB}{0,0,196}
\usepackage[colorlinks=true,linkcolor=darkblue,citecolor=darkblue,urlcolor=darkblue]{hyperref}
\newcommand*{\Njch}{\ensuremath{N_\mathrm{j,ch}}\xspace}
\newcommand*{\pT}{\ensuremath{p_\mathrm{T}}\xspace}

\newcommand*{\pTjet}{\ensuremath{p_\mathrm{T}^{\mathrm{ch,jet}}}\xspace}

\newcommand*{\GeVc}{\ensuremath{\mathrm{GeV}/c}\xspace}

\begin{document}
\title{Baryon enhancement in jets}

\author{Antonio Ortiz}%
 
\affiliation{%
 Instituto de Ciencias Nucleares, UNAM,
 Circuito exterior s/n Ciudad Universitaria, 04510 Mexico City, Mexico
}%
\author{R\'obert V\'ertesi} 
\email{vertesi.robert@wigner.hu}
\affiliation{HUN-REN Wigner Research Centre for Physics, P.O. Box 49, H-1525 Budapest}

\begin{abstract}

The enhancement of the baryon production relative to mesons in small-collision systems is considered a breakthrough result of the Large Hadron Collider since a similar effect in heavy-ion collisions is understood by invoking the formation of the strongly-interacting quark--gluon plasma.  In this paper, a baryon enhancement is reported for $p_{\rm T}^{\rm ch,\, jet}>15$\,GeV/$c$ jets produced in pp collisions at $\sqrt{s}=13$\,TeV simulated with PYTHIA8. The effect can be explained as a transition between quark-initiated jets (low jet multiplicities) to gluon-initiated jets (high jet multiplicities). The present result contributes to a deeper understanding of the multiplicity dependence of baryon enhancement across different collision systems.
 
\end{abstract}
\maketitle

The quark–gluon plasma (QGP), a deconfined state of quarks and gluons predicted to exist at extremely high temperatures and energy densities, is one of the most exciting subjects in high-energy nuclear physics. In heavy-ion collisions at the Relativistic Heavy Ion Collider (RHIC) and the Large Hadron Collider (LHC), compelling evidence for QGP formation has emerged through multiple observables~\cite{Busza:2018rrf}.  In particular, the light-flavor baryon-to-meson ratios as a function of $p_{\rm T}$ exhibit a bump-like structure at intermediate $p_{\rm T}$ (1.5--8.0\,GeV/$c$), which gradually increases with decreasing impact parameter (increasing centrality)~\cite{ALICE:2014juv,ALICE:2015dtd,ALICE:2019hno}. The effect is explained as an interplay between the collective motion of the system (heavier particles are boosted to higher momenta) and quark recombination in a dense QCD medium characterized by deconfinment~\cite{Fries:2004gw}. Another observation is that the $p_{\rm T}$-integrated proton-to-pion ratio exhibits a mild centrality dependence, the ratio decreases by around 20\% from the most peripheral to the most central Pb--Pb collisions. 

One of the breakthrough results of the ALICE experiment is the observation of a similar baryon-to-meson enhancement in small systems such as pp and p--Pb collisions~\cite{ALICE:2022wpn}. The enhancement is highly comparable to that observed in heavy-ion collisions. With increasing multiplicity, the enhancement intensifies and the peak position shifts to higher \pT. Moreover, the increase at intermediate momenta is accompanied by a corresponding depletion of the ratio at low-momenta, with the $p_{\rm T}$-integrated particle ratios exhibiting a modest decrease with the multiplicity increase. Although this behavior is consistent with the interpretation of hydrodynamic origin~\cite{Heinz:2019dbd}, other approaches considering string density effects can qualitatively explain the data~\cite{Bierlich:2014xba,OrtizVelasquez:2013ofg}. The question is now whether the baryon-to-meson ratio is a quantity uniquely sensitive to final state effects, and to what extent hydrodynamical evolution of the system and quark recombination can be constrained by fitting the models to baryon-to-meson ratios as a function of multiplicity. 
To challenge the final-state interpretation, this work presents predictions for intra-jet baryon-to-meson ratios as a function of $j_{\rm T}$ across different jet multiplicity classes. This study is motivated by recent results indicating that high QCD density effects can emerge in systems such as high-multiplicity jets, without necessarily requiring QGP formation~\cite{Baty:2021ugw,Vertesi:2024fwl}. The analysis is performed using PYTHIA8, which incorporates collective-like behavior driven by string density effects~\cite{Fischer:2016zzs}. 

In this paper, by analogy to heavy-ion collisions where most partonic interactions involve low momentum transfers, the low-$p_{\rm T}$ jet regime is explored. This novel approach considers a kinematic domain that is experimentally accessible with detectors like ALICE, which, owing to its unique particle-identification capabilities, has reported jet measurements featuring identified constituents across a broad transverse momentum range. To date, ALICE measurements of baryon-to-meson ratios in pp collisions have mainly been differential in jet $p_{\rm T}$~\cite{ALICE:2022ecr}. Additionally, a measurement of the longitudinal momentum fraction distributions of charmed hadrons within jets has also been reported~\cite{ALICE:2023jgm}.
Motivated by these considerations, this work investigates jet hadrochemistry as a function of jet multiplicity for $p_{\rm T}^{\rm ch,jet}>15$ GeV/$c$, aiming to establish a more differential reference for comparisons with inclusive analyses in pp collisions and to explore a potential connection with hadron production in e$^{+}$e$^{-}$ annihilation processes.


PYTHIA8 is a Monte Carlo event generator for high-energy particle collisions~\cite{Sjostrand:2007gs}. The generator incorporates multiple physics mechanisms including parton distribution functions, initial-state and final-state parton showers, multiparton interactions (MPI), and hadronization processes. The thermodynamical string fragmentation is an extension of the Lund string model in which hadron production from a breaking color string is treated using statistical–thermal principles, assigning Boltzmann-like weights to hadron species based on an effective temperature. This approach improves the description of particle yields and strangeness enhancement by treating string breakups as locally equilibrated systems~\cite{Fischer:2016zzs}. The model includes string density effects (close packing mechanism). Color reconnection model beyond leading color approximation (CR-BLC) is implemented through a string-length–minimization procedure constrained by SU(3) color rules, allowing junction formation and baryonic topologies. By dynamically rearranging color flow, it enhances baryon production and modifies final-state hadrochemistry while remaining consistent with perturbative color structure~\cite{Christiansen:2015yqa}. CR-BLC mode 2, which requires all dipoles involved in a reconnection to be causally connected, reproduces the enhanced number of charmed baryons relative to charmed mesons~\cite{ALICE:2021npz}.  

In this analysis, PYTHIA8 (version 8.309) was used for all simulations. Jets were generated using hard-QCD processes with MPI switched off, thereby suppressing the underlying event. However, the observables were also evaluated with MPI on with underlying-event subtraction, and the effects remained unchanged. Thermodynamical string fragmentation and CR-BLC mode 2 was employed, with parameters summarized in Refs.~\cite{Fischer:2016zzs,Christiansen:2015yqa}. For specific studies, hard production processes with only gluonic or quark partonic final states were also examined.

Jets were reconstructed from primary charged particles with a pseudorapidity $|\eta|<1$ and a transverse momentum $p_{\rm T}>0.15$ GeV/$c$, with the anti-$k_{\rm T}$ algorithm~\cite{Cacciari:2008gp}, employing the energy recombination scheme and a resolution parameter $R = 0.4$. Other resolution parameters were tested (0.5 and 0.8), and the effects remained unchanged.
In the analysis, the results are reported as a function of $j_{\rm T}$, which denotes the momentum component perpendicular to the jet axis.

The jet longitudinal momentum fraction carried by the leading charged particle,
$z_{\parallel}^{\mathrm{ch}}$, is defined as the component of the
leading charged particle momentum projected onto the jet axis, normalized
by the jet momentum,
\begin{equation}
z_{\parallel}^{\mathrm{ch}} =
\frac{{\bf p}_{\,\mathrm{ch}}^{\,\mathrm{lead}} \cdot \hat{\bf p}_{\mathrm{jet}}}
{|{\bf p}_{\mathrm{jet}}|},
\end{equation}
where ${\bf p}_{\,\mathrm{ch}}^{\,\mathrm{lead}}$ is the momentum of the
leading charged particle in the jet and $\hat{\bf p}_{\mathrm{jet}}$ is the
unit vector along the jet axis. This quantity represents the fraction of
the jet momentum carried by the leading charged particle along the jet
direction.

Charged jets with transverse momenta $p_{\rm T}^{\rm ch.jet} > 15$ GeV/$c$ were selected for the $j_{\rm T}$ analysis. For the $z^{\rm ch}_\parallel$ analysis, $7 < p_{\rm T}^{\rm ch.jet} < 15$ GeV/$c$ was chosen in order to match the kinematical criteria of existing data~\cite{ALICE:2023jgm}.

\begin{figure}[h!]
    \centering
\includegraphics[width=0.95\linewidth]{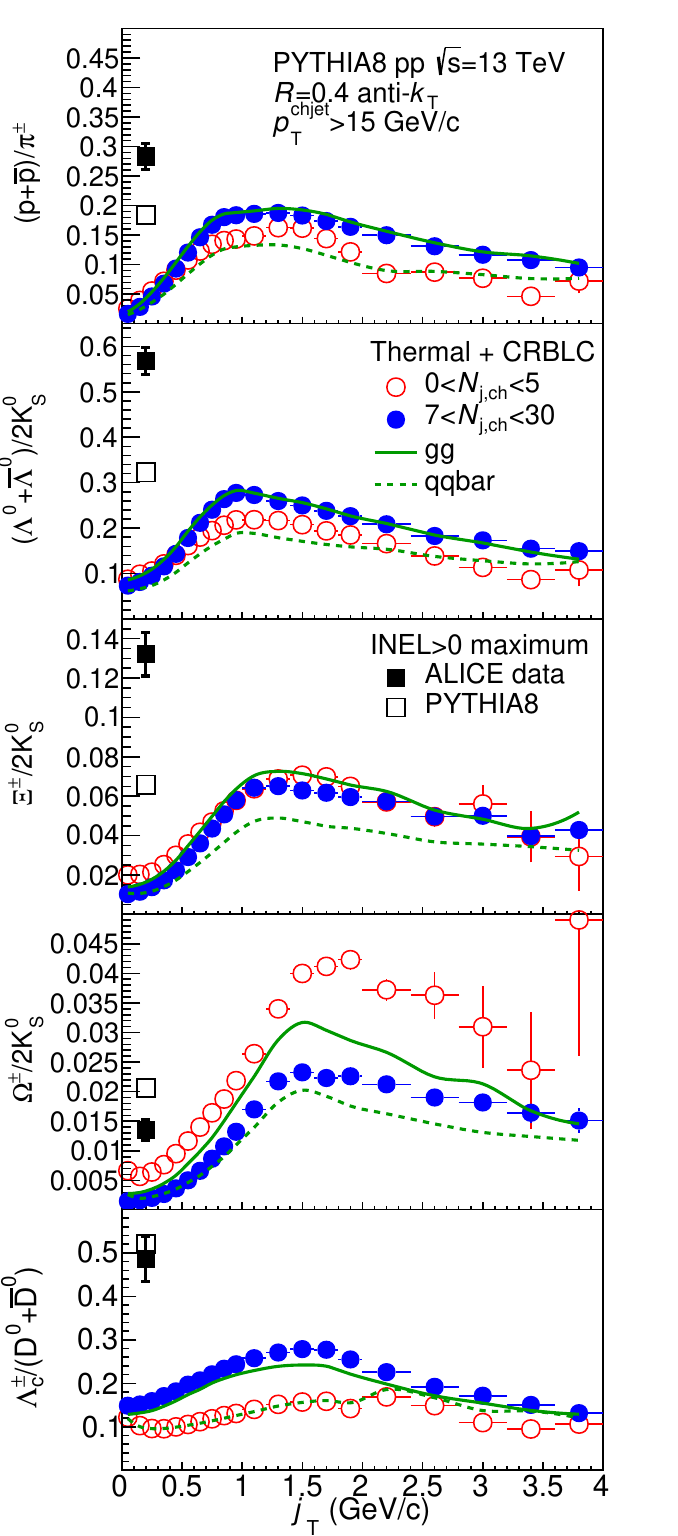}
\caption{Baryon-to-meson ratios as a function of $j_{\rm T}$ in pp collisions at $\sqrt{s}$ = 13 TeV, details about PYTHIA8 settings can be found in the text. High (low) multiplicity jets are displayed in red (blue) markers, while simulations for gluon (quark) jets are shown with a solid (dashed) line. The maximum values of minimum-bias data~\cite{ALICE:2020jsh,ALICE:2021npz} and simulations~\cite{Vertesi:2024fwl} are shown with solid and open squares, respectively.}
    \label{fig1}
\end{figure}

Figure~\ref{fig1} shows the baryon-to-meson yield ratios 
$\frac{\rm p + \bar{p}}{\rm \pi^{\pm}}$, 
$\frac{\rm \Lambda^{0} + \bar{\Lambda}^{0}}{2\,\rm K_{S}^{0}}$,
$\frac{\rm \Xi^{\pm}}{2\,\rm K_{S}^{0}}$,
$\frac{\rm \Omega^{\pm}}{2\,\rm K_{S}^{0}}$ and
$\frac{\rm \Lambda_{c}^{\pm}}{\rm D^{0} + \bar{D}^{0}}$ 
as a function of $j_{\rm T}$, for low and high charged-jet multiplicity classes 
($0<\Njch\leq 5$ and $7<\Njch\leq 30$, respectively). The particle ratios exhibit a bump structure at intermediate $j_{\rm T}$, in both charged-jet multiplicity classes. The higher multiplicity class shows a stronger baryon enhancement than the lower multiplicity class, except for multi-strange baryons ($\Xi$ and $\Omega$) where an opposite trend can be observed. A deviation of the behavior of $\Omega$ production relative to other hadrons has been also reported in the inclusive analysis as a function of multiplicity pp collisions due to phase-space constraints~\cite{Fischer:2016zzs}. Similar to the observations in pp and heavy-ion data, the enhancement is accompanied by a depletion at low $j_{\rm T}$. Figure~\ref{fig1} also shows simulations where only the gluon or the quark-producing hard processes are turned on. For comparison, the maximum values of the corresponding particle ratios in minimum-bias pp data~\cite{ALICE:2020jsh,ALICE:2021npz} and pp simulations~\cite{Vertesi:2024fwl} (at $\pT \approx 3$ \GeVc for protons over pions) are also plotted. All ratios, except for multi-strange baryons, exhibit a quark-gluon jet hierarchy:  ratios in low- and high-multiplicity jets are better described by quark- and gluon-initiated jets, respectively.

From LEP data, it is known that because of the color factor, gluons radiate more than quarks, giving a gluon jet a broader angular distribution and softer fragmentation than a quark jet~\cite{OPAL:1991ssr}. Moreover, gluon jets were found to exhibit higher total particle multiplicities and enhanced baryon production compared to quark jets~\cite{DELPHI:2000kzb}. Concerning the PYTHIA8 predictions, within the Lund fragmentation framework, the more complicated color/string environment (color topology) around a gluon can enhance baryon production relative to a simple quark endpoint. In such a model, the leading particle would more often be a baryon/antibaryon in a gluon jet than in a quark jet. The results from the simulations presented in this paper align with the expected different particle composition in quark and gluon jets~\cite{Andersson:1981ce}. This hypothesis was further tested by switching off color reconnection or the close packing mechanisms, the observed differences between quark and gluon jet particle ratios remained after such  variations. Concerning the observed jet-multiplicity dependence, we found that around 80\% of the low jet-multiplicity events were associated to quark jets, whereas, at high multiplicities, around 90\% was associated to gluon jets.

\begin{figure}[h!]
    \centering
\includegraphics[width=0.95\linewidth]{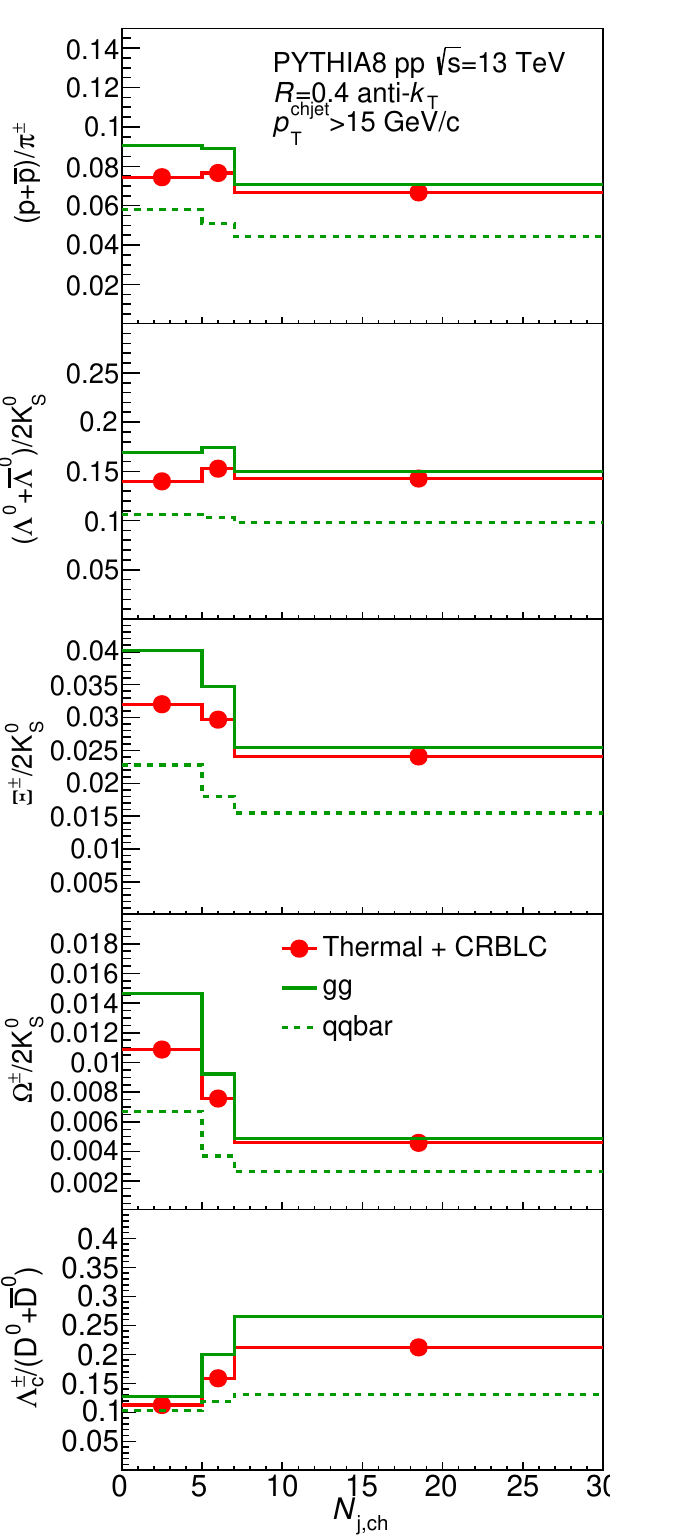}
    \caption{Baryon-to-meson ratios as a function of the jet charged-constituent multiplicity (red markers). Results are shown for pp collisions at $\sqrt{s}$ = 13 TeV, details about PYTHIA8 settings can be found in the text. Simulations for gluon (quark) jets are shown with a solid (dashed) line.}
    \label{fig2}
\end{figure}

The $j_{\rm T}$-integrated baryon-to-meson yield ratios as a function of the jet charged-constituent multiplicity can be seen in Fig.~\ref{fig2}.
Results for simulations where only the gluon or the quark-producing hard processes are turned on are also shown.
For the proton-to-pion and the $\Lambda^0$-to-K$_{S}^{0}$ ratios, the yield ratios are relatively constant over all three multiplicity classes. This is not true for the case where only quark and gluon processes are turned on, hinting that the quark--gluon composition of the jets change with charged-jet constituent multiplicity. Overall, the $j_{\rm T}$-integrated particle ratios are higher for gluon than for quark jets. Particularly, for the highest multiplicity class the proton-to-pion ratio is around 0.042 and 0.070 for quark and gluon jets, respectively. Such a multiplicity class matches the $\langle {\rm d}N_{\rm ch}/{\rm d}\eta\rangle$ ($|\eta|<0.8$) in minimum bias pp collisions at $\sqrt{s}=13$\,TeV. For this event class the $p_{\rm T}$-integrated proton-to-pion ratio is around 0.055~\cite{ALICE:2020nkc}, which is between the interval given by the prediction for quark and gluon jets. This observation indicates that minimum-bias pp data might have a larger gluon jet contribution than in simulations. The particle ratios involving multistrange hadrons exhibit a decreasing trend with increasing multiplicity probably due to phase-space constraints. Concerning the heavy-flavor particle ratio, the ratio increases with multiplicity; however, it is nearly flat for quark jets and multiplicity dependent for gluon jets. Remarkably, for the lowest multiplicity class the ratio is consistent with measurements in e$^{+}$e$^{-}$ annihilation processes~\cite{Gladilin:2014tba}. It is to be noted that PYTHIA8 with thermodynamical string fragmentation and CR-BLC is able to reproduce the \pT-differential intra-jet trends for both light, strange and multi-strange baryon-to-meson and baryon-to-baryon ratios~\cite{ALICE:2022ecr}.

\begin{figure*}
    \centering
\includegraphics[width=0.7\linewidth]{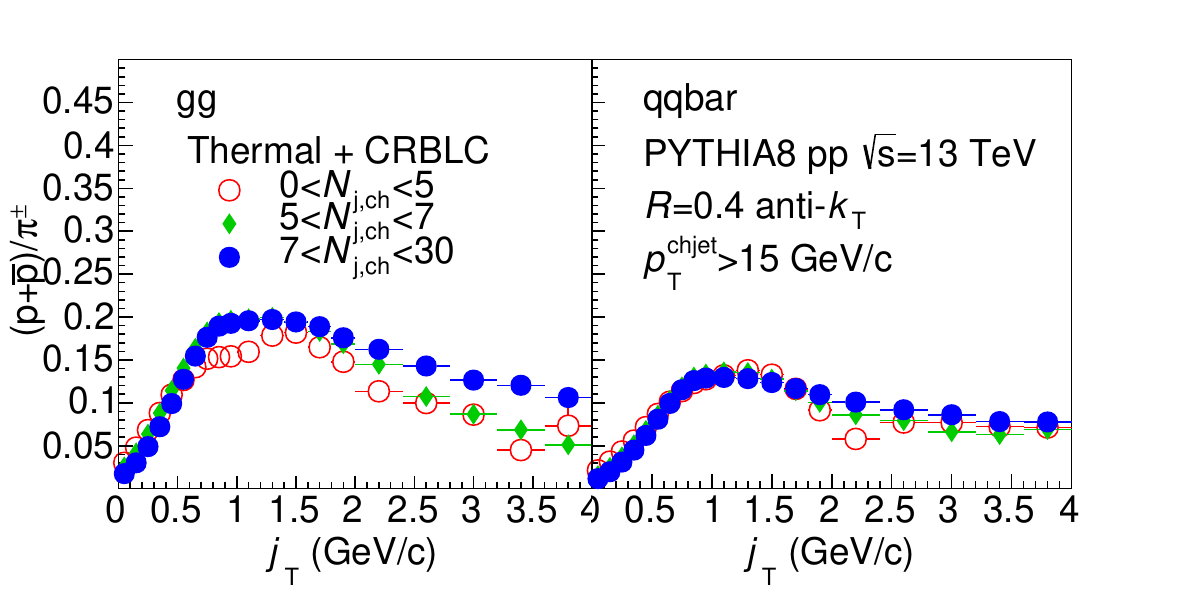}
    \caption{Proton-to-pion ratios as a function of $j_{\rm T}$ for gluon (left) and quark (right) jets in pp collisions at $\sqrt{s}$ = 13 TeV, details about PYTHIA8 settings can be found in the text. Different jet charged-constituent multiplicity intervals are shown.}
    \label{fig3}
\end{figure*}

Proton-to-pion ratios as a function of $j_{\rm T}$, separately for gluonic and quark-producing hard processes can be seen in Fig.~\ref{fig3} for different jet charged-constituent multiplicity intervals.   The particle ratios for quark jets exhibit little or no multiplicity dependence, whereas, a multiplicity dependence is seen in gluon jets. The same effects are observed even if string junctions or close packing mechanism is switched off.  

The ultra-small systems presented in this paper bridge the multiplicity gap between e$^{+}$e$^{-}$ and pp collisions. The evolution of the particle ratios with the jet multiplicity suggest that the multiplicity selection biases the sample from quark jets to gluon jets. The latter are more sensitive to fragmentation biases, as they tend to produce a higher number of charged particles. In this picture, the baryon-to-meson enhancement cannot be directly connected with the presence of hydrodynamical radial flow or quark recombination. The applicability of hydrodynamics in low-multiplicity pp collisions is already under debate. Observing the effects discussed here in experimental data would offer a promising opportunity to confront hydrodynamic interpretations with alternative scenarios based on initial-state effects.

It is worth noting that the results presented here are consistent with those reported, for example, in Ref.~\cite{OrtizVelasquez:2013ofg}. In pp collisions at LHC energies, partonic interactions occur at low $x$, favoring gluon-initiated jets. Therefore, pp collisions are already in the regimen dominated by gluon jets, which, according to the present results, are expected to exhibit a baryon-to-meson enhancement. Moreover, multiparton interactions in pp collisions produce multiple minijets (most of them gluon jets), and their parent partons can interact before  hadronization. In such a dense partonic environment, with a large string density, additional effects such as color reconnection may further amplify the baryon-to-meson enhancement.

\begin{figure}[h!]
    \centering
    \includegraphics[width=0.95\linewidth]{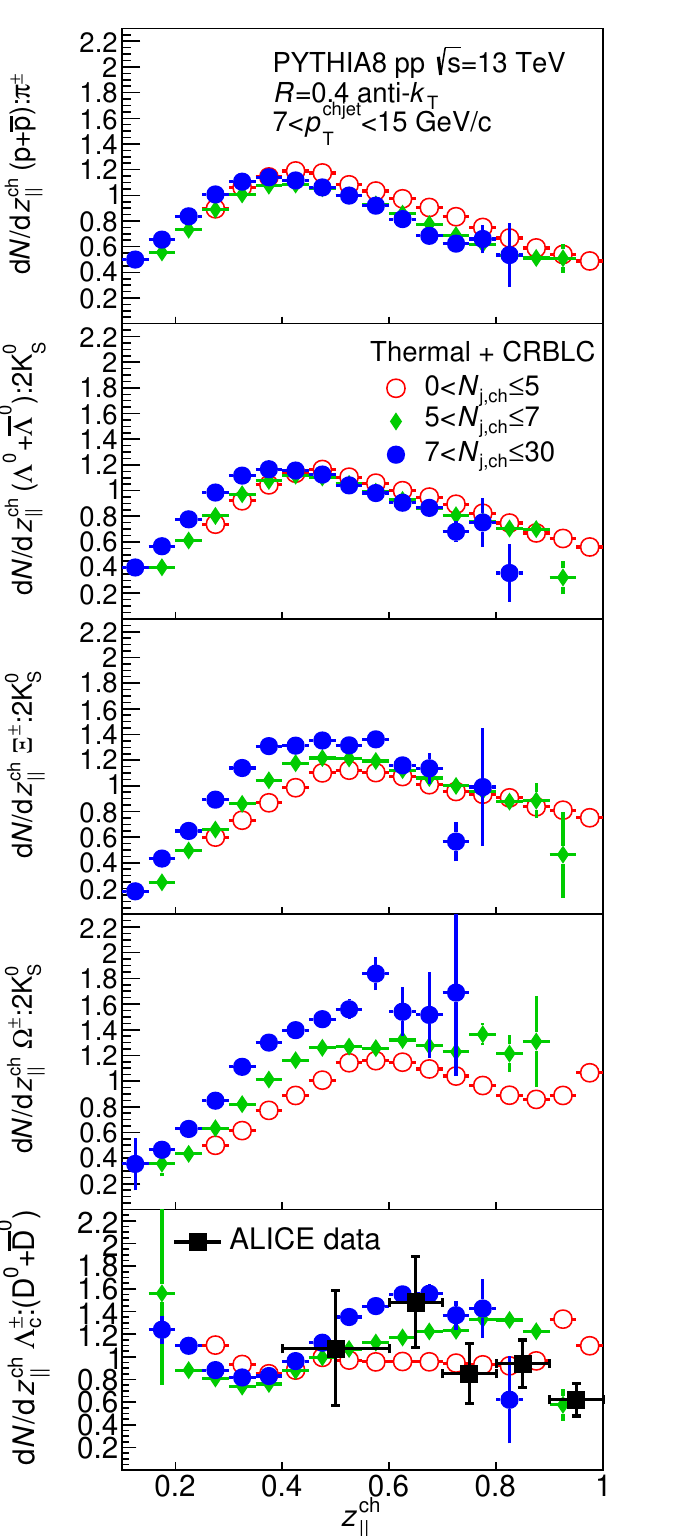}
    \caption{Ratios of $z^{\rm ch}_\parallel$ distributions for baryons over mesons, computed in different jet charged-constituent multiplicity intervals. Results are shown for pp collisions at $\sqrt{s}$ = 13 TeV, simulated with PYTHIA8 thermodynamical string fragmentation and color reconnection beyond leading color approximation settings, and no MPI. The heavy-flavor results are compared with data from ALICE~\cite{ALICE:2023jgm}.}
    \label{fig4}
\end{figure}

Figure~\ref{fig4} shows the ratios of $z^{\rm ch}_\parallel$ distributions for baryons over mesons, computed in different jet charged-constituent multiplicity intervals. For the case of the $\Lambda_c^0$ over ${\rm D}^0$ ratio, ALICE measurements are also shown for $7<\pTjet<15$ \GeVc and $3<p_{\rm T}^{\rm hadron}<15$ \GeVc. 
In the simulations, a pronounced multiplicity dependence is observed. In general, higher-multiplicity distributions are shifted to the left, meaning a softer fragmentation for baryons relative to the mesons. The charm data suggest that the fragmentation of charm quarks into charm baryons is softer than into charm mesons~\cite{ALICE:2023jgm}. This behavior is not reproduced by PYTHIA8 with thermodynamical string fragmentation and CR-BLC, although the simulations remain consistent with the measurements within current uncertainties. Future multiplicity-differential measurements with LHC Run~3 and Run 4 data will be essential to clarify these trends.

In summary, the baryon-to-meson ratios were analyzed in charged jets with $\pTjet>15$ \GeVc using PYTHIA8 simulations with thermodynamical string fragmentation and color reconnection beyond leading color approximation. 
The baryon-to-meson ratios as a function of $j_{\rm T}$ 
show a clear jet multiplicity dependence, particularly in gluon-dominated jets, whereas little or no multiplicity dependence is observed for quark-initiated jets. This behavior persists even when string junctions or the close packing mechanism are disabled, indicating that the effect is largely driven by the increasing fraction of gluon jets in high-multiplicity samples.

This interpretation is further supported by the charmed-hadron longitudinal momentum fraction ratios, which indicate softer fragmentation in high-multiplicity jets. This behavior is consistent with a larger contribution of gluon-initiated jets, known to yield softer fragmentation patterns and enhanced baryon production. 

These findings suggest that baryon-to-meson enhancement in small systems and jets may arise from parton-type and fragmentation biases rather than hydrodynamic flow or quark recombination. If confirmed experimentally, intra-jet particle ratios could provide a powerful tool to disentangle final-state collective effects from initial-state and fragmentation-driven mechanisms in high-energy collisions.

This work has been supported by DGAPA-UNAM PAPIIT No.\ IG100524 and PAPIME No.\ PE100426, as well as by the Hungarian National Research, Development and Innovation Office (NKFIH) under the contract numbers  
NKKP ADVANCED\_25-153456 and 2025-1.1.5-NEMZ\_KI-2025-00002, the Wigner Scientific Computing Laboratory (WSCLAB) and the HUN-REN Cloud.

\end{document}